\documentclass[11pt,twocolumn]{article}
\usepackage{graphicx}
\usepackage{amsmath}
\usepackage{hyperref}
\usepackage{booktabs}
\usepackage{tabularx}
\usepackage[backend=biber,style=numeric,sorting=none]{biblatex} 
\title{Critical Infrastructure Security: Penetration Testing and Exploit Development Perspectives}
\author{Papa Kobina Orleans-Bosomtwe \\ School of Computer Science, University of Guelph, Ontario, Canada}

\begin{document}

\maketitle

\begin{abstract}
\quad Critical infrastructure refers to essential physical and cyber systems vital to the functioning and stability of societies and economies. These systems include key sectors such as healthcare, energy, and water supply, which are crucial for societal and economic stability and are increasingly becoming prime targets for malicious actors, including state-sponsored hackers, seeking to disrupt national security and economic stability.
This paper reviews literature on critical infrastructure security, focusing on penetration testing and exploit development. It explores four main questions: the characteristics of critical infrastructure, the role and challenges of penetration testing, methodologies of exploit development, and the contribution of these practices to security and resilience.
The findings of this paper reveal inherent vulnerabilities in critical infrastructure and sophisticated threats posed by cyber adversaries. Penetration testing is highlighted as a vital tool for identifying and addressing security weaknesses, allowing organizations to fortify their defenses. Additionally, understanding exploit development helps anticipate and mitigate potential threats, leading to more robust security measures.
The review underscores the necessity of continuous and proactive security assessments, advocating for integrating penetration testing and exploit development into regular security protocols. By doing so, organizations can preemptively identify and mitigate risks, enhancing the overall resilience of critical infrastructure. The paper concludes by emphasizing the need for ongoing research and collaboration between the public and private sectors to develop innovative solutions for the evolving cyber threat landscape. This comprehensive review aims to provide a foundational understanding of critical infrastructure security and guide future research and practices.
\end{abstract}

\textbf{Keywords:} Cybersecurity, penetration testing, exploit development, critical infrastructure.

\section{Introduction}
\quad Critical infrastructure includes the physical and cyber systems and assets essential for the uninterrupted functioning of a nation’s society and economy \cite{Makrakis2021,1}. This includes healthcare, public health, energy, and information technology sectors. Over the years, advances in technology and information technology systems have improved many essential systems, making life easier for relevant users and operators. It is not shocking to discover critical infrastructure firms' assimilation and adoption of newer technological systems to improve efficiency and output.  The threats to technology increased with the progress and advancements over the years. Technological threats have become as sophisticated (sometimes more refined) as their counterparts. These threats can then be transferred to technologies used in critical infrastructure systems.\quad Cyber attackers, who pose said threats to technological systems, utilize various means to cause harm to systems. A survey in 2023 was conducted to find the primary cause of cyber-attacks Figure 1 encountered by companies in the United States, which showed unpatched vulnerabilities, attributing to 23\% of these causes. Penetration testing is basically an analysis of some aspects of a system \cite{Bishop2007,2}, with the discoveries used to improve the security and resilience of tested systems and networks by crafting new exploits or using existing ones to test systems further. Figure 2  shows the number of common IT security vulnerabilities and exposures worldwide (CVEs) from 2009 to 2024. This paper aims to assess the existing security and resilience of critical infrastructure using penetration testing and exploit development by answering 4 main research questions. These research questions are "What is critical infrastructure?", "What is penetration testing?", "What is exploit development?", and "How can penetration testing and exploit development improve critical infrastructure security and resilience?". These questions are being posed due to the information and understanding they can provide to readers of this paper. Answering these questions in the order they have been asked will help readers understand what critical infrastructure is, how important it is, and how the remaining concepts which are penetration testing, and exploit development tie in in the security of critical infrastructure. These questions will be answered by reviewing pre-existent studies around these key areas: critical infrastructure, penetration testing, and exploit development. 

\begin{figure}[h]
    \centering
    \includegraphics[width=\linewidth]{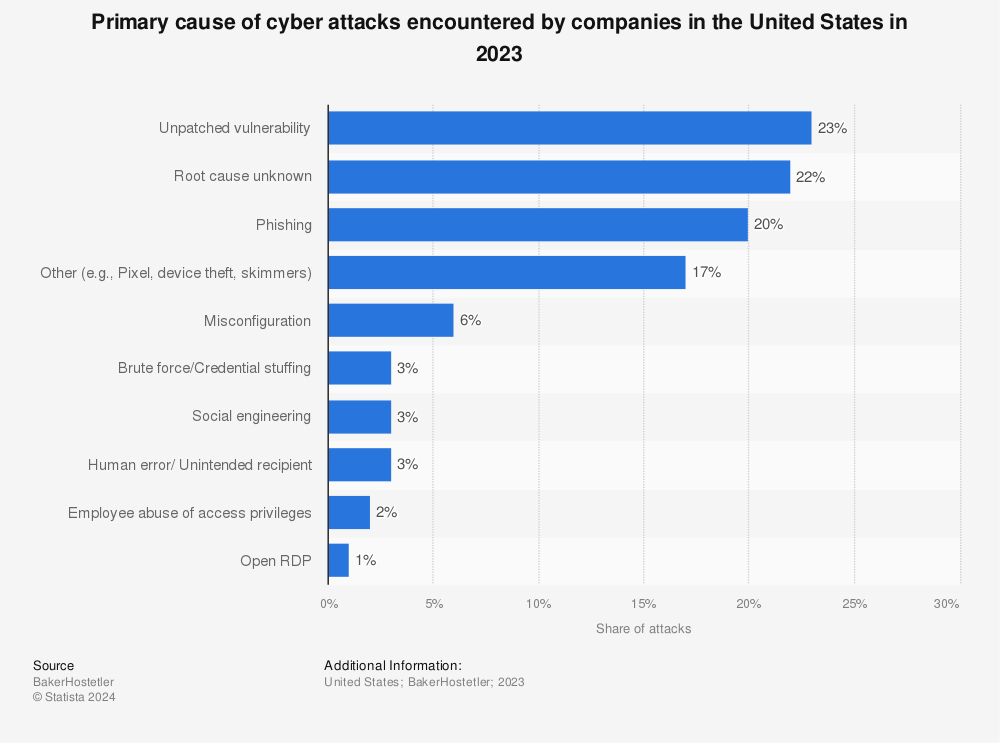}
    \caption{Primary causes of cyber-attacks in the US in 2023}
    \label{fig:figure1}
\end{figure}

\begin{figure}[h]
    \centering
    \includegraphics[width=\linewidth]{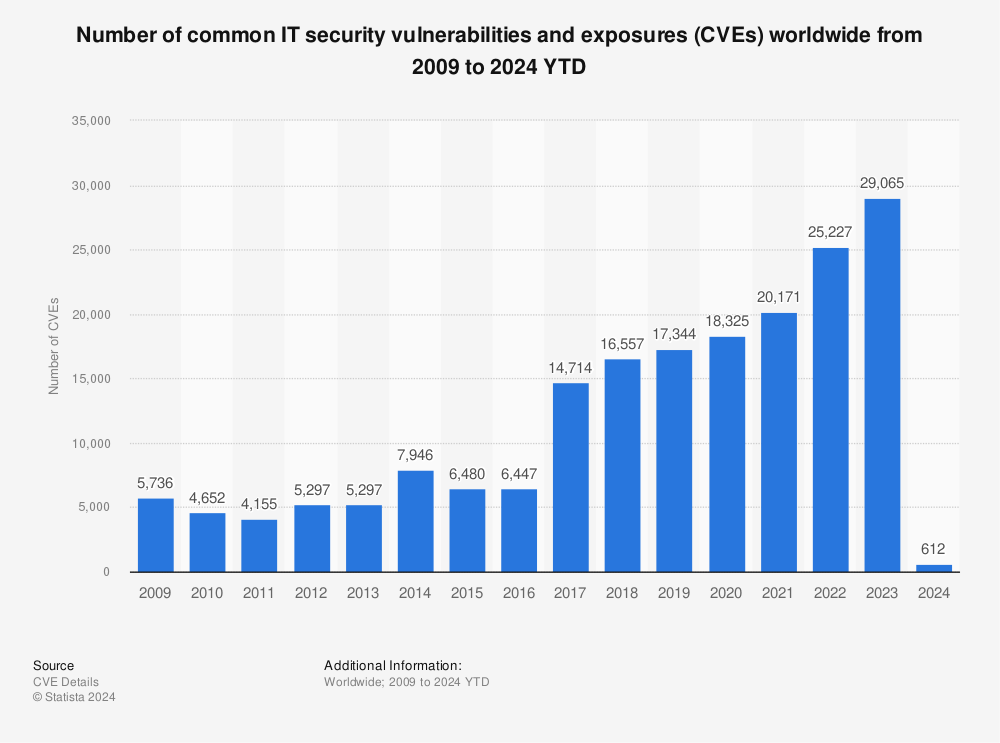}
    \caption{Number of common IT security vulnerabilities and exposures worldwide (CVEs) from 2009 to 2024}
    \label{fig:figure2}
\end{figure}

\section{Literature Review}
\quad Upon the first search of related artifacts, it was discovered that some articles related to critical infrastructure security from a cybersecurity standpoint. Dating to as recently as 2021, Makrakis et al. \cite{Makrakis2021} note the growing integration of modern information technology by critical infrastructure and industrial organizations into their existing operational technology (OT) architectures. They point out that the ever-increasing attack surfaces of these new modern technology integrations provide malicious attackers, owing to their complexity and modernity, which, in turn, also handicap the defenders of these systems. This paper then follows up with a survey of the most prevalent threats against industrial control systems and critical infrastructures at the time, exposing vulnerabilities in specific operational technology (OT) network protocols and devices. The authors also explore some malicious software (malware) that has targeted critical infrastructure in the past, highlighting how social engineering has been a significant attack vector for adversaries. 
\quad 2021 also saw Malatji et al. \cite{Malatji2021} reviewing critical infrastructure and cybersecurity in general, focusing on the growing interconnectivity between Enterprise Information Technology (IT) and Industrial Control Systems (ICS). The authors note that this growing connectivity also poses increasing dangers to operators and organizations regarding the new attack surfaces they present to malicious attackers. The paper also discusses the cybersecurity framework proposed by the National Institute of Standards and Technology (NIST) for critical infrastructure, noting its customization potential and inclusion of cloud capabilities and security in the modern world. 
\quad A paper by Ghafir et al. \cite{Ghafir2018} in 2018 points out threats to critical infrastructure, focusing on humans, classified as “non-computer experts.” The authors acknowledge current security awareness training platforms and tools and highlight their shortcomings. Using prior research, the authors note that the knowledge retention rate after completing a security awareness training session or campaign was low. Their paper proposes a “context-aware education tool” for security awareness, where training is related to the current business environment, with platform administrators able to monitor and track users' progress.
\quad The studies mentioned above all note the growing cybersecurity risks facing Critical Infrastructure (CI) ICS as they get updated and integrated with modern IT systems and networks. They do not, however, explore penetration testing and exploit development perspectives in improving the security of CI and ensuring their resilience. This paper aims to review further work done in this capacity to be a guide for future research activities \cite{1h}.

\section{Research Goals}
This paper aims to review existing studies on critical infrastructure security, penetration testing, and exploit development, and the utilization of penetration testing and exploit development to enhance critical infrastructure security and resilience.

\begin{table*}[!hbt]
    \begin{center}
    \begin{tabularx}{\textwidth}{|X|X|}
        \hline
        \textbf{Research Questions} & \textbf{Description} \\
        \hline
        \textbf{RQ1}: What is critical infrastructure? & A brief overview of critical infrastructure and what differentiated it from other structures put up to make it vital to a government and its population.  \\
        \hline
        \textbf{RQ2}: What is penetration testing? & An explanation of penetration testing, types of penetration testing, phases involved, common tools, and its challenges and limitations. \\
        \hline
        \textbf{RQ3}: What is exploit development? & This question aims to explore exploit development in its entirety, including types of exploits, some tools and techniques used, and ethics to be considered when developing exploits.   \\
        \hline
        \textbf{RQ4}: How can penetration testing and exploit development improve critical infrastructure security and resilience & A combination of all research questions above to finally understand critical infrastructure security from a penetration testing and exploit development standpoint.  \\
        \hline
    \end{tabularx}
    \caption{Research Questions}
    \label{tab:researchQuestions}
    \end{center}
\end{table*}

\begin{itemize}
    \item A total of 32 different articles and studies were discovered relating to critical infrastructure, its security, and penetration testing and exploit development. 
    \item After further review, 13 of the previous 32 articles and studies were selected for this paper. The selected 13 papers are carefully and comprehensively reviewed to carry out the purposes of this paper. 
\end{itemize}

The paper continues with a research methodology indicating how the selected journals and articles were found. The paper then presents all findings from the selected journals and discusses these findings. A conclusion will be made, and suggestions for possible future research efforts will be made.

\section{Research Methodology}
\subsection{Inclusion Criteria}
\quad Studies to be selected for this paper had to meet pre-defined criteria. This criterion includes an in-depth explanation and analysis of critical infrastructure security, penetration testing, exploit development, and relevant supporting background research. The studies must also be written within the last ten years and in English. All discovered studies on Google Scholar and other sources listed above were compared against the criteria defined in this section and Table 2. 

\subsection{Selection Results}
\quad The initial keyword searches yielded numerous results that had to be filtered through based on the inclusion criteria. This initial title/abstract screening yielded thirty studies relevant to the purposes of this paper. The thirty shortlisted studies subsequently underwent another screening phase (full text), closely following the inclusion criteria. Of the thirty initially found, only 13 papers were selected for final review in this paper.

\begin{table*}[!hbt]
    \begin{center}
    \begin{tabularx}{\textwidth}{|X|X|}
        \hline
        \textbf{Inclusion Criteria} & \textbf{Exclusion Criteria} \\
        \hline
        The paper must be written in English. & Paper not written in English. \\
        \hline
        The paper must discuss one, or more of the relevant topics. & Paper does not discuss critical infrastructure in a cybersecurity context. \\
        \hline
    \end{tabularx}
    \caption{Inclusion and Exclusion Criteria}
    \label{tab:selectionCriteria}
    \end{center}
\end{table*}

\subsection{Data Extraction}
\quad After going through the first and second screening steps, the papers that were finally selected had their data extracted. This data was extracted to evaluate the accuracy of information provided in the chosen texts. The data extraction process was first carried out on 3 out of the final 13 papers and was then applied to the remaining ten papers, with the data then getting categorized and stored for later use. The categories gained from the extracted data are as follows:
\begin{itemize}
    \item \textbf{Context data:} This includes all information about the purpose of the study. 
    \item \textbf{Qualitative data:} This includes the findings, proposals, and conclusions proposed by the studies' authors. 
\end{itemize}

\subsection{Significant Keyword Counts}
\quad In addition to the keywords utilized in searching for these papers, various other keywords were discovered in the selected texts. These keywords have been listed and compiled, with their given counts in Table 3. The aim of Table 3 is to accumulate the number of times selected keywords appeared in the 13 studies selected. Observing the keyword counts in Table 3, it can be noted that the prevailing theme across the selected studies is “cybersecurity,” with a count of 375. 

\begin{table*}[!hbt]
    \begin{center}
    \begin{tabular}{|c|c|}
        \hline
        \textbf{Keywords} & \textbf{Count} \\
        \hline
        critical infrastructure. & 213 \\
        \hline
        industrial control system & 80 \\
        \hline
        penetration testing/tests & 208 \\
        \hline
        cybersecurity & 375 \\
        \hline
        power system control & 6 \\
        \hline
        SCADA & 176 \\
        \hline
        exploit & 189 \\
        \hline
        threats & 145 \\
        \hline
    \end{tabular}
    \caption{Significant Keyword Counts}
    \label{tab:count-keywords}
    \end{center}
\end{table*}

\section{Findings}
\quad After carefully reading selected primary texts, relevant information related to this paper was extracted, including contextual and qualitative data. From the percentages in Figure 4, it can be determined that 27\% of the selected studies were concerned with cybersecurity. The next most prevalent themes in the selected studies were critical infrastructure and penetration testing, each with a 15\% distribution. Critical infrastructure being our main focus of this conversation and penetration testing being an aid in discovering some misconfigurations within critical infrastructure systems. At 14\%, the keyword “exploits” comes in as the fourth most discussed theme in the selected texts. Supervisory Control and Data Acquisition (SCADA) systems, which are typically used in industrial control devices and systems had a 13\% rate of discussion within the selected texts. Threats, mostly digital and within the cyber space have a 10\% discussion in the selected studies. Threats are the dangers that industrial control systems and critical infrastructure face. Industrial control systems usually found being used in Critical infrastructure had a 6\% distribution within the selected discussion papers. 
The selected studies all discussed the topics intended to meet their intended capacities. These studies were chosen because they sufficiently described and explained the relevant topics in this paper which are critical infrastructure, penetration testing, exploit development, and cybersecurity. A number of studies highlight the services and systems (IT) that are used in critical infrastructure and their gross misconfigurations. The importance of critical infrastructure penetration testing is also discussed, with some studies diving deeper into the importance of penetration testing and cybersecurity in critical infrastructure\cite{2h,3h}.

\begin{figure}[h]
    \centering
    \includegraphics[width=\linewidth]{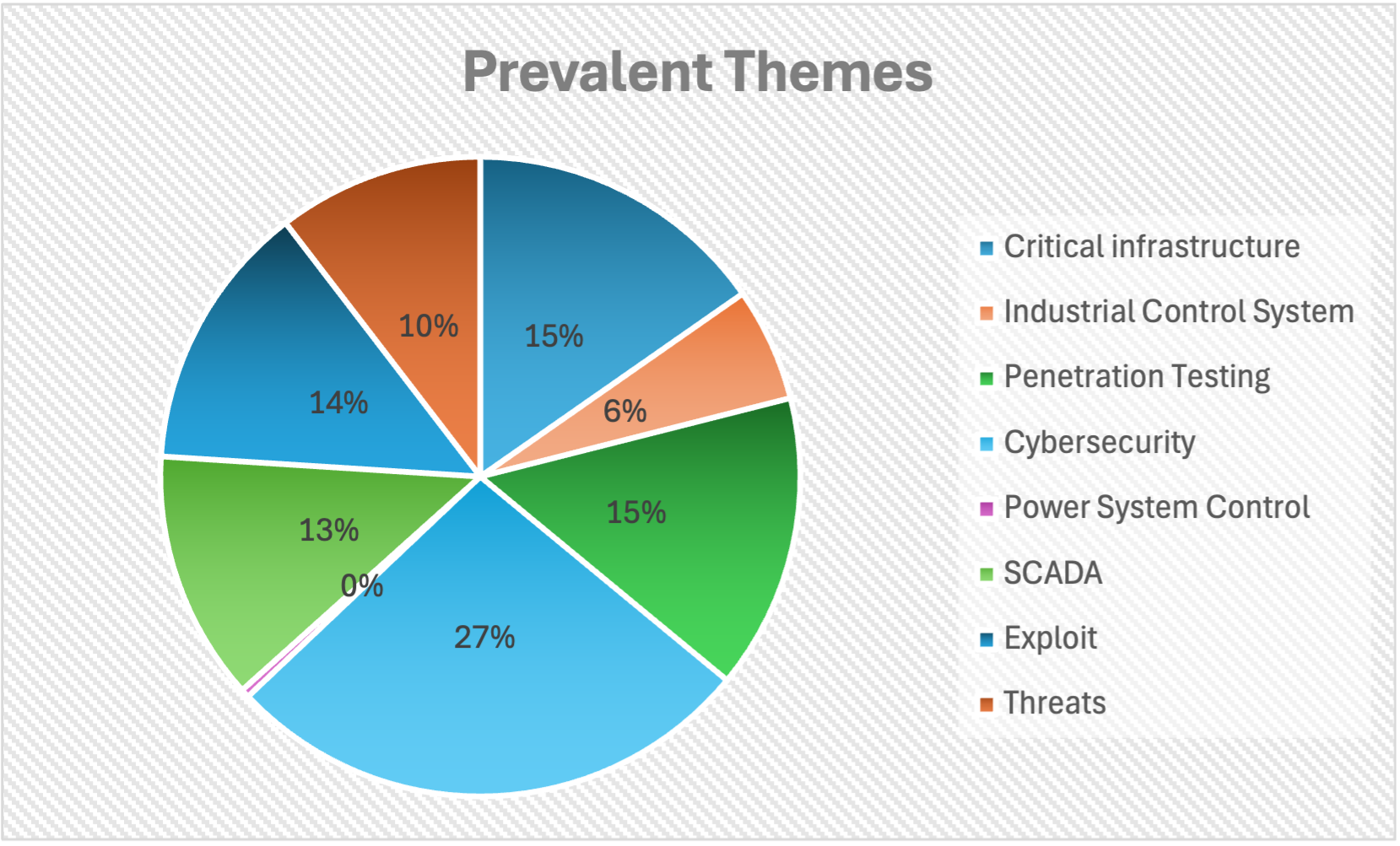}
    \caption{Distribution of Prevalent Themes}
    \label{fig:figure3}
\end{figure}

\section{Discussion}
The keyword searches conducted on the selected papers show a relevant interest in critical infrastructure security, penetration testing, and exploit development. The significance of critical infrastructure is well established in several of these studies, with a significant number also recognizing the growing risks these infrastructures face as they modernize their IT systems and other control systems. The papers have significant qualitative data regarding the chosen topics but need more practical solutions. Some valuable solutions offered are revisions of currently existing methodologies with some improvements. About penetration testing in these studies, the act is defined and well explained, with light also shed on the different types of penetration tests. Penetration testing tools were discussed, and a combination of these tools and other platforms was explored to simplify penetration testing. Addressing the critical infrastructure, a number of papers attempt to define CI and industrial control systems. The papers note that ICS were initially believed to be safe and free from possible cyber-attacks but increasing attacks on CI and ICS have changed this belief. IT and cybersecurity teams are now attempting to quantify the cyber risks CI and ICS are exposed to and are finding it difficult to do so. Multiple ICS and OT network environments have been discovered to be misconfigured and are easy to exploit. Penetration testing and red teaming activities have been shown to help discover said misconfigurations. Other researchers also propose using testbeds to discover cybersecurity vulnerabilities within ICS. Some authors of the selected studies attempt to develop penetration system specific to large networks found in CI by simulating complex human or automated attacks. Solutions proposed by authors include the integration of artificial intelligence and machine learning algorithms to detect and prevent potential cyber-attacks against ICS found within CI. The research also discovered that some definitions of Critical Infrastructure were inadequate, which could lead to some gaps in security as networks and systems for critical infrastructure were not properly categorized. CI must be well-defined, and assets, including systems and other products, must be correctly identified and labeled. The identified assets must be protected at all costs throughout their entire lifecycle. The remaining discoveries are tabulated in Figures 4-7 below.

\begin{figure}[h]
    \centering
    \includegraphics[width=\linewidth]{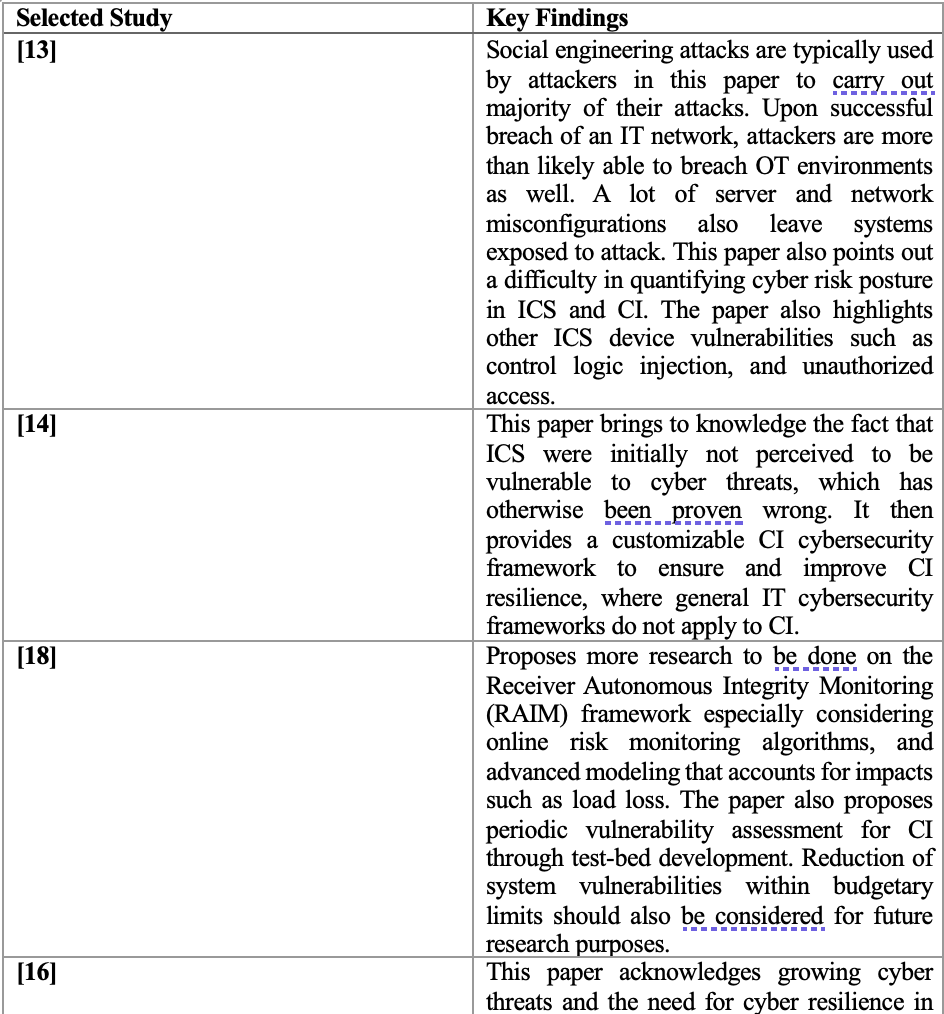}
    \caption{Key Findings}
    \label{fig:figure4}
\end{figure}

\begin{figure}[h]
    \centering
    \includegraphics[width=\linewidth]{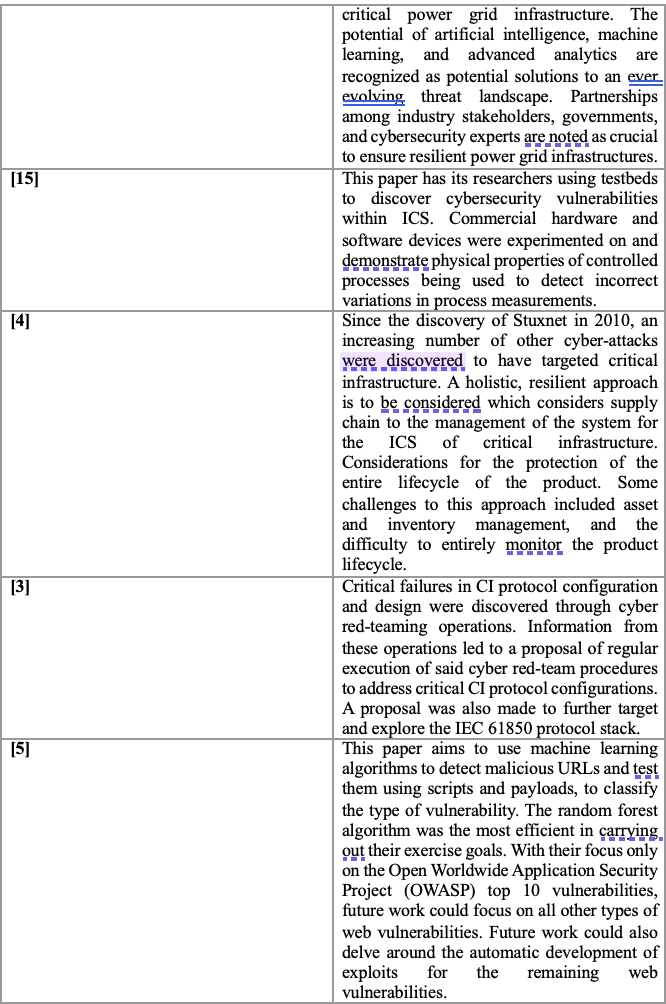}
    \caption{Key Findings continued}
    \label{fig:figure41}
\end{figure}

\begin{figure}[h]
    \centering
    \includegraphics[width=\linewidth]{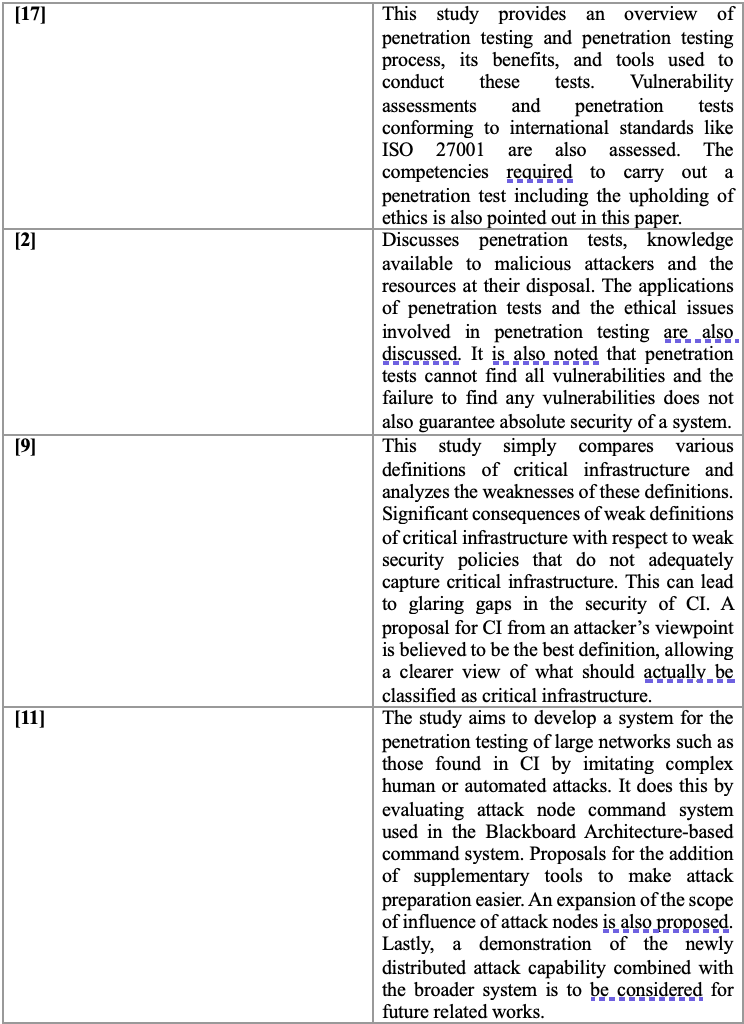}
    \caption{Key Findings continued}
    \label{fig:figure42}
\end{figure}

\begin{figure}[h]
    \centering
    \includegraphics[width=\linewidth]{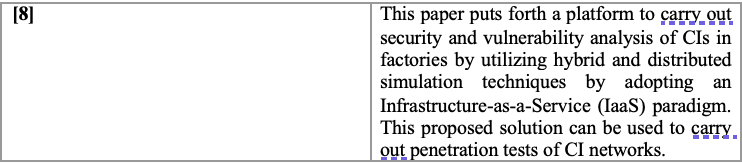}
    \caption{Key Findings continued}
    \label{fig:figure43}
\end{figure}

\subsection{RQ1. What is Critical Infrastructure?}
\quad Critical infrastructure can be defined as complex physical and cyber-based systems that form the lifeline of a modern society \cite{Ten2010,14}. Some sectors in Critical Infrastructure include transport, energy, food, water, finance, and health \cite{Ficco2017}. For various functions, industrial control systems are widely used in CI and are typically used to control different infrastructures like manufacturing \cite{Bologna2013}. The ICS architecture found in CI typically consists of 6 levels. These levels range from sensors providing sensing capabilities to a system, to manufacturing operations systems used to manage production at any CI site \cite{Makrakis2021,15}. The sentence before was added to paint a picture of just how complex ICS and CI can get and how all the systems found within the ICS archcitecture require security on varying levels to protect them from cyber attacks.

\subsection{RQ2. What is Penetration Testing?}
\quad Penetration testing is a structured process to test an organization's computing base, which includes hardware, software, and people \cite{Shebli2018}. The adopted frameworks differ in the number of penetration testing phases they offer. For example, the National Institute of Standards and Technology Special Publication (NIST SP) 800-115 has four stages, whereas the penetration testing execution standard (PTES) has seven. A penetration test's primary objectives are identifying vulnerabilities, reducing risks, and guaranteeing compliance by figuring out how a malevolent actor may enter the target environment, navigate it, and steal data from it. In penetration testing, three main testing methods are typically used, the first being black box. In black box testing, the team has no information about the tested target \cite{Shebli2018}. White box testing has testers provided with all information about the test target, while gray box testing has testers provided with partial information \cite{Shebli2018}. Several tools like Nmap, Metasploit, and Nessus are used in various stages of the penetration testing process. 

\subsection{RQ3. What is Exploit Development?}

\quad Exploit development involves identifying vulnerabilities in applications and software and determining how to gain control of a system \cite{Chadha2022}. When discussing exploit development, there are two broad categories under which all exploits fall: known and unknown (zero-day) exploits \cite{Chadha2022}. Both malicious actors and security teams can exploit developments to their advantage. Security team members or researchers can use exploit development to learn more about security flaws within systems or networks, including possible critical failures and the consequences of attackers exploiting these vulnerabilities. The information obtained from this exercise can help security team members adequately patch or fix the discovered vulnerabilities to prevent facing the consequences. This information can also be shared with relevant parties to ensure more systems affected by these vulnerabilities are secured.

\quad While the traditional methods of exploit development are critical, recent advancements in automated exploit generation tools and machine learning techniques have significantly enhanced the speed and efficacy of this process. Automated exploit generation tools, such as those leveraging symbolic execution and fuzz testing, allow for rapid identification and exploitation of vulnerabilities. These tools can systematically explore numerous execution paths in software to uncover hidden bugs and security flaws that might be missed by manual testing \cite{avgerinos2014automatic}.

\quad Recent advancements in automated exploit generation have significantly enhanced the speed and effectiveness of exploit development. Automated exploit generation typically involves automatically discovering paths in a program that trigger vulnerabilities, thereby creating exploits \cite{wang2023aaheg}. Tools like AAHEG exemplify the integration of automation in this field. AAHEG utilizes symbolic execution to analyze and detect potential heap-related vulnerabilities in sourcee code, develops an exploit abstract syntax tree, then selects exploitable methods. These methods are then tested and the final exploit is produced.

\quad Incorporating these technologies into exploit development can provide security teams with powerful tools to proactively identify and address vulnerabilities before malicious actors can exploit them. This proactive approach is essential for maintaining robust security measures against the continuously evolving cyber threat landscape.

\subsection{RQ4. How can penetration testing and exploit development improve critical infrastructure security and resilience?}

\quad The general issue with critical infrastructure security and resilience is that cybersecurity is primarily perceived in the context of enterprise IT systems. Industrial Control Systems (ICS) were mistakenly considered impervious to cyber-attacks \cite{Malatji2021}. Other issues relating to critical infrastructure include the old age of the systems and technologies used. Legacy systems and older components typically found in critical infrastructure system architectures pose significant problems for CI security teams \cite{Parshivlyuk2024}. These older systems typically lack modern security features to detect or counter contemporary cyber threats and attacks. Penetration testing and exploit development, when correctly carried out, especially on networks by the relevant parties, can expose vulnerabilities and possible threats to critical infrastructure.

\quad ICS used in CI consist of various levels and are very complex. This requires that security operators have deep understandings of the multiple tiers within ICS. Penetration testing, therefore, plays a critical role in ensuring the security and resilience of these systems by addressing vulnerabilities at each level \cite{Makrakis2021} of the architecture:

\begin{itemize}
\item \textbf{Level 0 - Sensors, motors, and instruments:} Penetration testing at this level involves assessing the security of devices that provide sensing capabilities to the ICS. These components are often targeted because they are fundamental to the physical operation of industrial processes. Penetration tests should check for vulnerabilities such as weak authentication mechanisms, insecure communication protocols, and potential physical tampering points \cite{11}.
\item \textbf{Level 1 - Devices including PLCs:} Programmable Logic Controllers (PLCs) are critical as they provide sensory and monitoring control over physical processes. Comprehensive penetration tests should focus on identifying flaws in PLC firmware, misconfigurations, and unsecured connections that could be exploited to disrupt operations or cause physical damage.
\item \textbf{Level 2 - Control systems:} This level includes Engineering Workstations that supervise physical processes. Penetration testing here involves evaluating the security of control systems, ensuring that they are not vulnerable to attacks that could lead to unauthorized changes in operational settings or shutdowns \cite{14}.
\item \textbf{Level 3 - Plant-wide production workflow systems:} Systems at this level, such as file servers and Microsoft Active Directory, are tested for vulnerabilities that could allow attackers to gain control over the workflow management systems. This includes examining network security, access controls, and potential insider threats.
\item \textbf{Level 4 - IT-related activity systems:} Systems like application servers and ERP systems are crucial for overseeing IT-related activities. Comprehensive penetration tests will assess these systems for vulnerabilities in web applications, databases, and network interfaces, ensuring they are robust against exploits that could impact the broader enterprise operations.
\item \textbf{Level 5 - Enterprise network:} The enterprise network encompasses both internal and external networks of the organization, used for production and resource data exchange. Penetration testing should focus on identifying and mitigating risks associated with network perimeter security, remote access vulnerabilities, and potential data breaches that could propagate to lower levels \cite{4h,5h}.
\end{itemize}

\quad Emphasis should be placed on network penetration testing because numerous incidents attest that attackers can easily penetrate Operational Technology (OT) environments after breaking into IT networks \cite{Makrakis2021}. A comprehensive penetration testing strategy that addresses each layer of the ICS architecture ensures that vulnerabilities are identified and mitigated across the entire system, thereby improving the security and resilience of critical infrastructure.

\quad Recent advancements in exploit development, particularly in automated exploit generation (AEG), can significantly enhance penetration testing and overall security of CI and ICS. Automated tools like AAHEG (Automatic Advanced Heap Exploit Generation) leverage symbolic execution and abstract syntax trees to automatically identify and exploit heap-related vulnerabilities without requiring source code, effectively bypassing various protection mechanisms \cite{wang2023aaheg}.

\quad Integrating these advanced technologies into penetration testing enables quicker and more efficient identification and mitigation of vulnerabilities, ensuring that security measures are robust and up-to-date against evolving cyber threats. This proactive approach is crucial for maintaining the security and resilience of critical infrastructure and industrial control systems.

\section{Conclusion and Future Work}
\quad Critical Infrastructure is vital to the daily operation and working of a society; any disruptions or outages to the services they provide cannot be afforded. Owing to their importance, critical infrastructure security should be taken seriously, especially in cyberspace. The growing attraction of cyber attackers to critical infrastructure is undeniable, and action should be taken. Penetration testing to discover the vulnerabilities in essential systems of critical infrastructure, coupled with exploit development to understand the criticality of these vulnerabilities, will spur security teams into immediate action to fix and patch these vulnerabilities and corresponding systems to prevent exploitation by malicious attackers. The research has shown the benefits of penetration testing and exploit development and pointed out, as earlier stated, how necessary critical infrastructure is. 
\quad For future research, I propose that researchers investigate social engineering and securing the human part of any IT system. Social engineering is currently the most common way of committing cybercrimes through the intrusion and infection of computer systems \cite{KlimburgWitjes2021}. Hypothetically speaking, even if the systems and network of critical infrastructure manage to be perfectly configured with no potential breach avenues, the one remaining threat will remain the human aspect of these systems. 
\quad Another risk aspect of the human part of any IT system is insider threats. Insider threats are malicious acts carried out by authorized persons, which may cause detrimental implications for the digital and physical assets of an organization \cite{Alsowail2022}. More research should be conducted in this respect for mitigation strategies and the role penetration testing can play. 
\quad The final area of suggestion for future research will be artificial intelligence and machine learning. Machine learning and artificial intelligence could play essential roles in automated penetration testing and exploit development to benefit the cybersecurity of critical infrastructure systems.

\nocite{*}

\printbibliography

@Article{Ghafir2018,
  author    = {Ghafir, Ibrahim and Saleem, Jibran and Hammoudeh, Mohammad and Faour, Hanan and Prenosil, Vaclav and Jaf, Sardar and Jabbar, Sohail and Baker, Thar},
  journal   = {The Journal of Supercomputing},
  title     = {Security threats to critical infrastructure: The human factor},
  year      = {2018},
  month     = {3},
  number    = {10},
  pages     = {4986-5002},
  volume    = {74},
  doi       = {10.1007/s11227-018-2337-2},
  publisher = {Springer Science and Business Media LLC},
}

@Article{Chesne2024,
  author    = {Chesne, A},
  journal   = {Journal of Sound and Vibration},
  title     = {Indirect boundary force measurements in beam-like structures using a derivative estimator},
  year      = {2024},
  month     = {6},
  number    = {24},
  pages     = {6438-6452},
  volume    = {333},
  doi       = {10.1016/j.jsv.2014.07.026},
  publisher = {Elsevier BV},
}

@Article{DalalanaBertoglio2017,
  author    = {Dalalana Bertoglio, Daniel and Zorzo, Avelino},
  journal   = {Journal of the Brazilian Computer Society},
  title     = {Overview and open issues on penetration test},
  year      = {2017},
  month     = {2},
  number    = {1},
  volume    = {23},
  doi       = {10.1186/s13173-017-0051-1},
  publisher = {Sociedade Brasileira de Computacao - SB},
}

@Article{KlimburgWitjes2021,
  author    = {Klimburg-Witjes, Nina and Wentland, Alexander},
  journal   = {Science, Technology, \& Human Values},
  title     = {Hacking Humans? Social Engineering and the Construction of the “Deficient User” in Cybersecurity Discourses},
  year      = {2021},
  month     = {2},
  number    = {6},
  pages     = {1316-1339},
  volume    = {46},
  doi       = {10.1177/0162243921992844},
  publisher = {SAGE Publications},
}

@Article{Alsowail2022,
  author    = {Alsowail, Rakan and Al-Shehari, Taher},
  journal   = {PeerJ Computer Science},
  title     = {Techniques and countermeasures for preventing insider threats},
  year      = {2022},
  month     = {4},
  pages     = {e938},
  volume    = {8},
  doi       = {10.7717/peerj-cs.938},
  publisher = {PeerJ},
}

@Article{Makrakis2021,
  author    = {Makrakis, Georgios and Kolias, Constantinos and Kambourakis, Georgios and Rieger, Craig and Benjamin, Jacob},
  journal   = {IEEE Access},
  title     = {Industrial and Critical Infrastructure Security: Technical Analysis of Real-Life Security Incidents},
  year      = {2021},
  pages     = {165295-165325},
  volume    = {9},
  doi       = {10.1109/access.2021.3133348},
  publisher = {Institute of Electrical and Electronics Engineers (IEEE)},
}

@Article{Malatji2021,
  author    = {Malatji, Masike and Marnewick, Annlizé and Von Solms, Suné},
  journal   = {Information \& Computer Security},
  title     = {Cybersecurity capabilities for critical infrastructure resilience},
  year      = {2021},
  month     = {10},
  number    = {2},
  pages     = {255-279},
  volume    = {30},
  doi       = {10.1108/ics-06-2021-0091},
  publisher = {Emerald},
}

@Article{Ten2010,
  author    = {Ten, Chee-Wooi and Manimaran, Govindarasu and Liu, Chen-Ching},
  journal   = {IEEE Transactions on Systems, Man, and Cybernetics - Part A: Systems and Humans},
  title     = {Cybersecurity for Critical Infrastructures: Attack and Defense Modeling},
  year      = {2010},
  month     = {7},
  number    = {4},
  pages     = {853-865},
  volume    = {40},
  doi       = {10.1109/tsmca.2010.2048028},
  publisher = {Institute of Electrical and Electronics Engineers (IEEE)},
}

@Article{Parshivlyuk2024,
  author  = {Parshivlyuk, S and Panchenko, K},
  journal = {Eduzone: International Peer Reviewed/Refereed Multidisciplinary Journal},
  title   = {Cyber Threats and Resilience in Power Grid Infrastructures: Assessing Vulnerabilities and Countermeasures},
  year    = {2024},
  number  = {1},
  volume  = {13},
  pages   = {1},
}

@Article{Morris2011,
  author    = {Morris, Thomas and Srivastava, Anurag and Reaves, Bradley and Gao, Wei and Pavurapu, Kalyan and Reddi, Ram},
  journal   = {International Journal of Critical Infrastructure Protection},
  title     = {A control system testbed to validate critical infrastructure protection concepts},
  year      = {2011},
  month     = {8},
  number    = {2},
  pages     = {88-103},
  volume    = {4},
  doi       = {10.1016/j.ijcip.2011.06.005},
  publisher = {Elsevier BV},
}

@InBook{Bologna2013,
  author    = {Bologna, Sandro and Fasani, Alessandro and Martellini, Maurizio},
  editor    = {M. Martellini},
  pages     = {57-72},
  publisher = {Springer International Publishing},
  title     = {Cyber Security and Resilience of Industrial Control Systems and Critical Infrastructures},
  year      = {2013},
  booktitle = {Cyber Security},
  doi       = {10.1007/978-3-319-02279-6_6},
}

@InProceedings{Blumbergs2019,
  author    = {Blumbergs, Bernhards},
  booktitle = {Proceedings of the 5th International Conference on Information Systems Security and Privacy},
  title     = {Remote Exploit Development for Cyber Red Team Computer Network Operations Targeting Industrial Control Systems},
  year      = {2019},
  pages     = {88--99},
  publisher = {SCITEPRESS - Science and Technology Publications},
  doi       = {10.5220/0007310300880099},
}

@InBook{Chadha2022,
  author    = {Chadha, Rosy and Shalom, G and Anand, Vivek and Goel, Agam},
  pages     = {1--7},
  publisher = {IEEE},
  title     = {A Study on Exploit Development},
  year      = {2022},
  month     = {11},
  booktitle = {2022 7th International Conference on Computing, Communication and Security (ICCCS)},
  doi       = {10.1109/icccs55188.2022.10079387},
}

@InBook{Shebli2018,
  author    = {Shebli, Hessa and Beheshti, Babak},
  pages     = {1--7},
  publisher = {IEEE},
  title     = {A study on penetration testing process and tools},
  year      = {2018},
  month     = {5},
  booktitle = {2018 IEEE Long Island Systems, Applications and Technology Conference (LISAT)},
  doi       = {10.1109/lisat.2018.8378035},
}

@Article{Bishop2007,
  author    = {Bishop, Matt},
  journal   = {IEEE Security \& Privacy Magazine},
  title     = {About Penetration Testing},
  year      = {2007},
  month     = {11},
  number    = {6},
  pages     = {84-87},
  volume    = {5},
  doi       = {10.1109/msp.2007.159},
  publisher = {Institute of Electrical and Electronics Engineers (IEEE)},
}

@Article{Gallais2017,
  author    = {Gallais, C and Filiol, E},
  journal   = {The Journal of Trauma: Injury, Infection, and Critical Care},
  title     = {Intrathecal administration of tetanus antitoxin and corticosteroids in treatment of tetanus},
  year      = {2017},
  number    = {11},
  pages     = {893},
  volume    = {17},
  doi       = {10.1097/00005373-197711000-00029},
  publisher = {Ovid Technologies (Wolters Kluwer Health)},
}

@Article{Hance2022,
  author    = {Hance, Jack and Milbrath, Jordan and Ross, Noah and Straub, Jeremy},
  journal   = {Computers},
  title     = {Distributed Attack Deployment Capability for Modern Automated Penetration Testing},
  year      = {2022},
  month     = {2},
  number    = {3},
  pages     = {33},
  volume    = {11},
  doi       = {10.3390/computers11030033},
  publisher = {MDPI AG},
}

@Article{Ficco2017,
  author    = {Ficco, Massimo and Choraś, Michał and Kozik, Rafał},
  journal   = {Journal of Computational Science},
  title     = {Simulation platform for cyber-security and vulnerability analysis of critical infrastructures},
  year      = {2017},
  month     = {9},
  pages     = {179-186},
  volume    = {22},
  doi       = {10.1016/j.jocs.2017.03.025},
  publisher = {Elsevier BV},
}

@Article{avgerinos2014automatic,
  author    = {Avgerinos, Thanassis and Cha, Sang Kil and Rebert, Adam and Schwartz, Edward J. and Woo, Maverick and Brumley, David},
  journal   = {Communications of the ACM},
  title     = {Automatic exploit generation},
  year      = {2014},
  volume    = {57},
  number    = {2},
  pages     = {74--84},
  doi       = {10.1145/2560217.2560219},
  publisher = {ACM},
}

@Article{wang2023aaheg,
  author    = {Wang, Yu and Zhang, Yipeng and Li, Zhoujun},
  journal   = {Symmetry},
  title     = {AAHEG: Automatic Advanced Heap Exploit Generation Based on Abstract Syntax Tree},
  year      = {2023},
  volume    = {15},
  number    = {12},
  pages     = {Article 12},
  doi       = {10.3390/sym15122197},
  publisher = {MDPI},
}

@article{1,
  title={An energy-efficient SDN controller architecture for IoT networks with blockchain-based security},
  author={Yazdinejad, Abbas and Parizi, Reza M and Dehghantanha, Ali and Zhang, Qi and Choo, Kim-Kwang Raymond},
  journal={IEEE Transactions on Services Computing},
  volume={13},
  number={4},
  pages={625--638},
  year={2020},
  publisher={IEEE}
}

@article{2,
  title={Enabling drones in the internet of things with decentralized blockchain-based security},
  author={Yazdinejad, Abbas and Parizi, Reza M and Dehghantanha, Ali and Karimipour, Hadis and Srivastava, Gautam and Aledhari, Mohammed},
  journal={IEEE Internet of Things Journal},
  volume={8},
  number={8},
  pages={6406--6415},
  year={2020},
  publisher={IEEE}
}

@article{3,
  title={P4-to-blockchain: A secure blockchain-enabled packet parser for software defined networking},
  author={Yazdinejad, Abbas and Parizi, Reza M and Dehghantanha, Ali and Choo, Kim-Kwang Raymond},
  journal={Computers \& Security},
  volume={88},
  pages={101629},
  year={2020},
  publisher={Elsevier}
}

@article{4,
  title={Block hunter: Federated learning for cyber threat hunting in blockchain-based iiot networks},
  author={Yazdinejad, Abbas and Dehghantanha, Ali and Parizi, Reza M and Hammoudeh, Mohammad and Karimipour, Hadis and Srivastava, Gautam},
  journal={IEEE Transactions on Industrial Informatics},
  volume={18},
  number={11},
  pages={8356--8366},
  year={2022},
  publisher={IEEE}
}

@article{5,
  title={Secure intelligent fuzzy blockchain framework: Effective threat detection in iot networks},
  author={Yazdinejad, Abbas and Dehghantanha, Ali and Parizi, Reza M and Srivastava, Gautam and Karimipour, Hadis},
  journal={Computers in Industry},
  volume={144},
  pages={103801},
  year={2023},
  publisher={Elsevier}
}

@article{6,
  title={Efficient design and hardware implementation of the OpenFlow v1. 3 Switch on the Virtex-6 FPGA ML605},
  author={Yazdinejad, Abbas and Bohlooli, Ali and Jamshidi, Kamal},
  journal={The Journal of Supercomputing},
  volume={74},
  pages={1299--1320},
  year={2018},
  publisher={Springer}
}

@article{7,
  title={A high-performance framework for a network programmable packet processor using P4 and FPGA},
  author={Yazdinejad, Abbas and Parizi, Reza M and Bohlooli, Ali and Dehghantanha, Ali and Choo, Kim-Kwang Raymond},
  journal={Journal of Network and Computer Applications},
  volume={156},
  pages={102564},
  year={2020},
  publisher={Elsevier}
}

@article{1h,
  title={A generalizable deep neural network method for detecting attacks in industrial cyber-physical systems},
  author={Sakhnini, Jacob and Karimipour, Hadis and Dehghantanha, Ali and Yazdinejad, Abbas and Gadekallu, Thippa Reddy and Victor, Nancy and Islam, Anik},
  journal={IEEE Systems Journal},
  volume={17},
  number={4},
  pages={5152--5160},
  year={2023},
  publisher={IEEE}
}

@inproceedings{8,
  title={A machine learning-based sdn controller framework for drone management},
  author={Yazdinejad, Abbas and Rabieinejad, Elnaz and Dehghantanha, Ali and Parizi, Reza M and Srivastava, Gautam},
  booktitle={2021 IEEE Globecom Workshops (GC Wkshps)},
  pages={1--6},
  year={2021},
  organization={IEEE}
}

@article{9,
  title={AP2FL: Auditable privacy-preserving federated learning framework for electronics in healthcare},
  author={Yazdinejad, Abbas and Dehghantanha, Ali and Srivastava, Gautam},
  journal={IEEE Transactions on Consumer Electronics},
  year={2023},
  publisher={IEEE}
}

@article{10,
  title={Hybrid privacy preserving federated learning against irregular users in next-generation Internet of Things},
  author={Yazdinejad, Abbas and Dehghantanha, Ali and Srivastava, Gautam and Karimipour, Hadis and Parizi, Reza M},
  journal={Journal of Systems Architecture},
  volume={148},
  pages={103088},
  year={2024},
  publisher={Elsevier}
}

@article{11,
  title={A Robust Privacy-Preserving Federated Learning Model Against Model Poisoning Attacks},
  author={Yazdinejad, Abbas and Dehghantanha, Ali and Karimipour, Hadis and Srivastava, Gautam and Parizi, Reza M},
  journal={IEEE Transactions on Information Forensics and Security},
  year={2024},
  publisher={IEEE}
}

@article{14,
  title={An ensemble deep learning model for cyber threat hunting in industrial internet of things},
  author={Yazdinejad, Abbas and Kazemi, Mostafa and Parizi, Reza M and Dehghantanha, Ali and Karimipour, Hadis},
  journal={Digital Communications and Networks},
  volume={9},
  number={1},
  pages={101--110},
  year={2023},
  publisher={Elsevier}
}

@article{15,
  title={Accurate threat hunting in industrial internet of things edge devices},
  author={Yazdinejad, Abbas and Zolfaghari, Behrouz and Dehghantanha, Ali and Karimipour, Hadis and Srivastava, Gautam and Parizi, Reza M},
  journal={Digital Communications and Networks},
  volume={9},
  number={5},
  pages={1123--1130},
  year={2023},
  publisher={Elsevier}
}

@phdthesis{12,
  title={Secure and private ml-based cybersecurity framework for industrial internet of things (iiot)},
  author={Yazdinejad, Abbas},
  year={2024},
  school={University of Guelph}
}

@article{2h,
  title={Federated quantum-based privacy-preserving threat detection model for consumer internet of things},
  author={Namakshenas, Danyal and Yazdinejad, Abbas and Dehghantanha, Ali and Srivastava, Gautam},
  journal={IEEE Transactions on Consumer Electronics},
  year={2024},
  publisher={IEEE}
}

@article{3h,
  title={The dichotomy of cloud and iot: Cloud-assisted iot from a security perspective},
  author={Zolfaghari, Behrouz and Yazdinejad, Abbas and Dehghantanha, Ali and Krzciok, Jacob and Bibak, Khodakhast},
  journal={arXiv preprint arXiv:2207.01590},
  year={2022}
}

@article{4h,
  title={A Federated Learning Approach for Multi-stage Threat Analysis in Advanced Persistent Threat Campaigns},
  author={Nelles, Florian and Yazdinejad, Abbas and Dehghantanha, Ali and Parizi, Reza M and Srivastava, Gautam},
  journal={arXiv preprint arXiv:2406.13186},
  year={2024}
}

@article{h5,
  title={Security considerations for virtual reality systems},
  author={Viswanathan, Karthik and Yazdinejad, Abbas},
  journal={arXiv preprint arXiv:2201.02563},
  year={2022}
}

\end{document}